\documentclass{IEEEtran}

\usepackage{graphicx}
\usepackage{url}
\usepackage{float}
\usepackage{tabularx}
\usepackage{multirow}
\usepackage{listings}
\usepackage{color}
\usepackage{xspace}
\usepackage{paralist}
\usepackage{array}
\usepackage{cite}
\usepackage{fixltx2e}
\usepackage{soul}
\usepackage{amsmath}

 \definecolor{mauve}{rgb}{0.58,0,0.82}
 \definecolor{mygreen}{rgb}{0,0.6,0}

\newif\ifdraft
\draftfalse
\ifdraft
 \newcommand{\jhanote}[1]{ \textcolor{red}  {***SJ:#1}}
 \newcommand{\mynote}[1]{ {\textcolor{blue} { ***VB: #1 }}}
 \newcommand{\atnote}[1]{}
 \newcommand{\rmnote}[1]{}
 \newcommand{\shrink}[1]{}
\else
 \newcommand{\jhanote}[1]{}
 \newcommand{\mynote}[1]{}
 \newcommand{\atnote}[1]{}
 \newcommand{\rmnote}[1]{}
 \newcommand{\shrink}[1]{}
\fi

\newcommand{\up}{\vspace*{-1em}}

\newcommand{\rp}{RADICAL-Pilot\xspace}
\newcommand{\enmd}{Ensemble toolkit\xspace}

\newcommand\FramedImage[2][]{%
  \setlength\fboxsep{.5pt}% change according to needs
  \setlength\fboxrule{1pt}%
  \noindent\fbox{%
    \begin{minipage}[c][\dimexpr.16\textheight-1.5\fboxrule-2\fboxsep\relax][c]{\dimexpr.33\textwidth-1.5\fboxrule-2\fboxsep\relax}
    \centering
    \includegraphics[width=\textwidth, height=0.65\textwidth]{#2}
  \end{minipage}}%
}

\begin{document}
%\setcopyright{acmcopyright}
%\conferenceinfo{PASC16}{June 8--10, 2016, Lausanne, Switzerland}
%\CopyrightYear{2016} 

\title{Ensemble Toolkit: Scalable and Flexible Execution of Ensembles of Tasks}

%\numberofauthors{4}

\author{

\IEEEauthorblockN{
Vivekanandan Balasubramanian,
Antons Treikalis,
Ole Weidner\IEEEauthorrefmark{1},
Shantenu Jha\IEEEauthorrefmark{2}}

\IEEEauthorblockA{Department of Electrical and Computer Engineering, Rutgers University, Piscataway, New Jersey 08854}

\IEEEauthorblockA{
 \IEEEauthorrefmark{1}Work done when at Rutgers}

\IEEEauthorblockA{
 \IEEEauthorrefmark{2}Corresponding Author}

}

\maketitle

\begin{abstract} 
There are many science applications that require scalable task-level parallelism, support for flexible execution and coupling of ensembles of simulations.  Most high-performance system software and middleware, however, are designed to support the execution and optimization of single tasks.  Motivated by the missing capabilities of these computing systems and the increasing importance of task-level parallelism, we introduce the Ensemble toolkit which has the following application development features: (i) abstractions that enable the expression of ensembles as primary entities, and (ii) support for ensemble-based execution patterns that capture the majority of application scenarios. Ensemble toolkit uses a scalable pilot-based runtime system that decouples workload execution and resource management details from the expression of the application, and enables the efficient and dynamic execution of ensembles on heterogeneous computing resources. We investigate three execution patterns and characterize the scalability and overhead of Ensemble toolkit for these patterns.  We investigate scaling properties for up to  O(1000) concurrent ensembles and O(1000) cores and find linear weak and strong scaling behaviour.
\end{abstract}

\atnote{I think linear scaling is ambiguous: in some communities,
    scaling is defined in terms of parallel efficiency, and in this case I don't
    think ENMD shows linear scaling for all patterns} 

%\keywords{Keyword 1; Keyword 2; Keyword 3;}

\section{Introduction}
\label{sec:intro}

Many scientific applications in the field of molecular sciences, computational biology \cite{laughton2009coco,preto2014fast,cheatham2015impact}, astrophysics \cite{sirko2005initial}, weather forecasting \cite{wan2014short,bauer2015quiet}, bioinformatics \cite{martin2010rnnotator} are increasingly reliant on the ability to run ensemble-based methods to make scientific progress.  This is true for applications that are both net producers of data, as well as aggregate consumers of data. Computationally, an ensemble is comprised of multiple concurrent units of execution, which are referred to as {\it tasks}. Tasks may refer to simulations, analysis or any independent computational process that needs to be executed.  Different ensemble-based application vary in the degree of coupling between tasks, dependency between stages of tasks, as well as the level of heterogeneity across tasks.

In spite of the apparent simplicity of running ensemble-based applications, the scalable and flexible execution of a large and collective set of tasks is non-trivial. The requirements of ensemble-based applications are more complex than simple parameter sweeps: they require varying degrees of coupling between the ensemble members, such as global synchronization, temporally varying pairwise interactions. In addition, all ensemble members must successfully complete as the collective properties of the entire ensemble are often computed.

% Given the complexity and the many degrees-of-freedom, it is not surprising that scalable and flexible tools for ensemble-based applications are conspicuous by their absence. 

There are many functional, performance and usability challenges that currently prevent ensemble-based applications from exploiting the full power of high-performance computing (HPC) systems. The situation is exacerbated by the fact that HPC software eco-system has traditionally been designed and optimized with large, monolithic applications.  There is an absence of tools designed and implemented with ensembles as the fundamental unit.  Hitherto, ensemble-based applications have either been retrofitted into one of many {\it general purpose} workflow systems, or wrapped in specifically constructed scripts that address some, but rarely all of the requirements. The former can in principle support ensemble-based applications as a {\it special} case, but most workflow systems have their design and performance tuned to managing complex inter-dependencies of tasks, which is not the %distinguishing characteristic or
dominant requirement of ensemble-based applications. The scripting-based approach presents a significant burden on the user, especially, when it comes to running large ensembles in a fault-tolerant way on heterogeneous resources or in developing sophisticated ensemble-based applications.

%\jhanote{By the way what is the difference between fault tolerance and reliable and resilience?? typical exam question!}

In response to these challenges and the growing importance of ensemble-based applications, we design and implement \enmd with the following features to meet the requirements of ensemble-based applications: (i) abstractions for the expression of an ensemble of tasks, (ii) support for ensemble-based execution patterns, (iii) decoupling the execution and management of these patterns from their expression and, (iv) a runtime system for efficient execution of tasks and provides flexible resource utilization capabilities over a range of HPC platforms.

%keeping the new software footprint small and reusing  

\jhanote{Does it do previous things better? Or new things that were not possible?}  \mynote{previous things better I guess, ppl have been running ensemle-based applications (with much hardships) on HPCs}\jhanote{this should come out somewhat explicitly in the writing} \jhanote{By the way the last sentence in following paragraph is not entirely consistent with the viewpoint that its better and not new..}
 
The design and implementation of \enmd advances both systems and applications. While being novel in design and concept, its implementation builds upon well-established abstractions and significant reuse of existing software components.  Although the original motivation of \enmd arose from the needs of the bio-molecular science~\cite{extasy-2016} and geoscience applications, the concepts, components and use of \enmd is agnostic to the application domain.  \enmd fills a missing gap in the repertoire of production-grade HPC tools.
% designed to capture the {\it sweet-spot} in the space of control, flexibility and usability, 

%In this paper, we present the Ensemble MD (EnMD) toolkit starting with information on some of the terminology used in the paper in section 2. In section 3, we discuss the objectives that we aim to address using the EnMD toolkit followed by the design of the toolkit. We provide a brief introduction to the API exposed to the user and conclude the section with a description of the advantages/disadvantages of using this toolkit. In section 4, we discuss the design of the experiments to characterize the toolkit followed by experiments to test the scalability of the toolkit. In section 5, we present the implications of the results obtained and discuss avenues of optimizations that can be made to the toolkit. We conclude the paper in section 6 with future development ideas.  

After a brief discussion in Section 2 of related tools and research, in Section 3, we present the objectives, design, and implementation of \enmd. In Section 4, we perform validation and scalability experiments using different workloads. We conclude in Section 5 with a discussion of the key features and aspects of \enmd.

%comparison of the key features  of \enmd with other similar tools.

\section{Related Work}
\label{sec:related}

\shrink{none have been designed with ensembles as the fundamental entity and to support flexible ways to execute them.}

Many tools have been developed to run workloads on HPC resources, however to the best of our knowledge, there are no best practices for ensemble-based applications, nor are there tools that embody those practises.  For example, replica-exchange and other ensemble-based enhanced sampling algorithms in molecular science applications~\cite{preto2014fast, laughton2009coco, sugita1999replica} use approaches that depend on the number of tasks, % (typically O(100)-O(1000)),
coupling and size of the system being studied. Similarly, in geosciences, ensemble-based simulations are used to study dispersion of atmospheric pollutants, effects of contaminant release~\cite{cervone2008risk}, hurricane prediction~\cite{hamill2010future}; existing solutions handle the coordination, orchestration and execution of ensembles differently.

% and many others.  Within the aforementioned geoscience applications, 

% due to the illusion of providing complete control and flexibility. 

\jhanote{The is a reptition in some ways from the introduction. Should we use this real estate for something else?}\jhanote{Vivek -- the following two sentences are still a repitition of what was said in the introduction, with little new information..}\mynote{took another crack} \jhanote{the following two sentences will ``doom'' the paper: it is too general, too dismissive and factually incorrect}\mynote{took another crack}\jhanote{Mostly fine.}

In addition to the parameter sweep tools (e.g., Nimrod~\cite{nimrod}) and scripting methods discussed earlier, multiple workflow systems~\cite{taylor2014workflows} have been designed for HPC systems.  Typically these are geared towards exploiting concurrency across complex task dependencies and typically use custom languages that increase the effort from the user-end. Systems such as Pegasus\cite{deelman2004pegasus} and DAGMan\cite{team2005dagman} convert given workloads into directed acyclic graphs (DAGs). DAGMan simply schedules the jobs as per the DAG where each edge of the DAG specifies the order of precedence and primarily meant for static workloads. Pegasus
, primarily expressed in virtual data language, maps workflows onto select execution sites by creating resource specific DAG. Workflow systems such as Swift\cite{swift} do not rely on the creation of a DAG but require an explicit mention of the order of execution of the various tasks in a multi-stage application.

There are other dataflow oriented tools such as Copernicus~\cite{copernicus}, Ruffus\cite{Ruffus}, Snakemake\cite{koster2012snakemake} and COSMOS \cite{gafni2014cosmos}. Both Ruffus and COSMOS provide a python library to construct the user application whereas Snakemake provides a separate workflow definition language. Snakemake and COSMOS convert their workload into DAGs using tool dependencies and file naming conventions and enable execution on common HPC. Ruffus mainly concentrates on expression \& automation of conventional pipeline based workloads.

To the best of our knowledge, few, if any workflow systems have been designed for ensemble-based applications and their execution patterns from first principles. Tigres~\cite{hendrix2016tigres}, a recent library, for workflow composition provides ``programming templates'' to compose computational or data pipelines. By combining these templates it is possible to compose many ensemble-based applications. YAWL~\cite{van2005yawl} provides a workflow language that enables application composition by supporting a list of patterns that were observed to be common in workflows.

Some tools can be used for ensemble-based applications with additional customization, but the extent and level at which additional customization is required varies across tools. One of the motivations of this paper is to design components, using a general-purpose language, to provide domain-independent functionality and thus minimize this ``last mile'' effort.

% \jhanote{I don't understand what ``stage'' at which this effort become required varies means?}\mynote{fixed} 

% \jhanote{I can't figure out connection between domain independence and last mile?. Needs tightening}{I made minor changes, but please take a go at it}. \jhanote{In general, I also think that the real estate can be used more effectively Points of possible comparison: patterns based approach in workflows; }

\rmnote{The user must explicitly call commands to transfer files, submit programs as jobs for execution and implement control flow; this limits portability across HPC platforms} \shrink{Other web-service based tools %Some of the simple scripting approaches have also been interfaced with web-services based tools to manage job submission,
  such as Longbow\cite{longbow} and GROWL\cite{growl}. %However, both Longbow and GROWL
  are limited in the application supported as Longbow only supports Amber~\cite{pearlman1995amber}, CHARMM~\cite{brooks2009charmm}, Gromacs~\cite{van2005gromacs}, LAMMPS~\cite{plimpton2007lammps} or NAMD~\cite{phillips2005scalable} based simulations and GROWL is limited to C and C++ based applications.}

\section{Ensemble Toolkit}
\label{sec:enmdtk}

%\jhanote{This section in general needs (i) significant removal of repetitions, (ii) a structure that has subsections that mirror the flow as follows: requirements to design to implementation}
%\mynote{made another attempt}
% \input{enmd/intro}
% \input{enmd/design}
% \input{enmd/rp}
% \input{enmd/interface}

%\jhanote{not good style to begin a section and then immediately a new subsection, i.e., some intervening text is good practise}

% In this section we discuss the current status of methods to run task-parallel science applications on HPC systems and 

We provide a semi-informal discussion of the requirements of ensemble-based applications and methods to motivate the design and implementation of \enmd.

\subsection{Requirements}

% These applications consist of a large number of tasks that vary in the degree-of-coupling between the tasks as measured by degree of frequency of interaction, volume of information exchange between them. 

% There are many science and engineering pplications ranging from molecular sciences to astrophysics that take an ensemble-based approach to their solutions. 

\shrink{There are a number of requirements that must be met in order to support the wide range of ensemble-based applications.}  

Ensemble-based applications vary in the type of coupling between tasks, the frequency and volume of exchange between these tasks, and the executable software (``kernel'') of each ensemble member.  Applications developers require simple and uniform approaches for developing diverse ensemble-based applications without compromising scale and the ability to use heterogeneous resources.

%\jhanote{this is not a requirement: The two-step MapReduce can be generalized to an iterative simulation of swarms~\cite{preto2014fast} with an intermediate analysis that is more sophisticated than a simple reduce.}

% for a range of applications comprised of multiple distinct but related tasks.
%One of the primary challenges is the need 

From a systems perspective, there is a need to provide efficient and flexible resource management over a diverse range of resource types. Often the total resource requirements of the ensemble of tasks are much greater than the resources instantaneously available. This can be due to the lack of resources, long queue wait times or upper limits enforced by scheduling policies. In either case, it is important to decouple the total resources required, from the resources utilized (or available) at any instant of time.  This is not easily available on HPC systems and points to the need for a special-purpose runtime system.

\subsection{Design}

%\jhanote{please aim for directed writing, as opposed to ``meandering'': \enmd was designed to be a scalable execution framework for multiple tasks with varying coupling on different HPC platforms.  Given the wide range of ensemble-based applications, the toolkit aspect derives from the design objective of providing building blocks so as to make it extensible and flexible.  In addition to functional and performance requirements, \enmd was designed with software engineering best practices so as to be production quality and grade.}

Three primary design objectives of \enmd are:
\begin{figure}
\centering
    \includegraphics[width=0.45\textwidth]{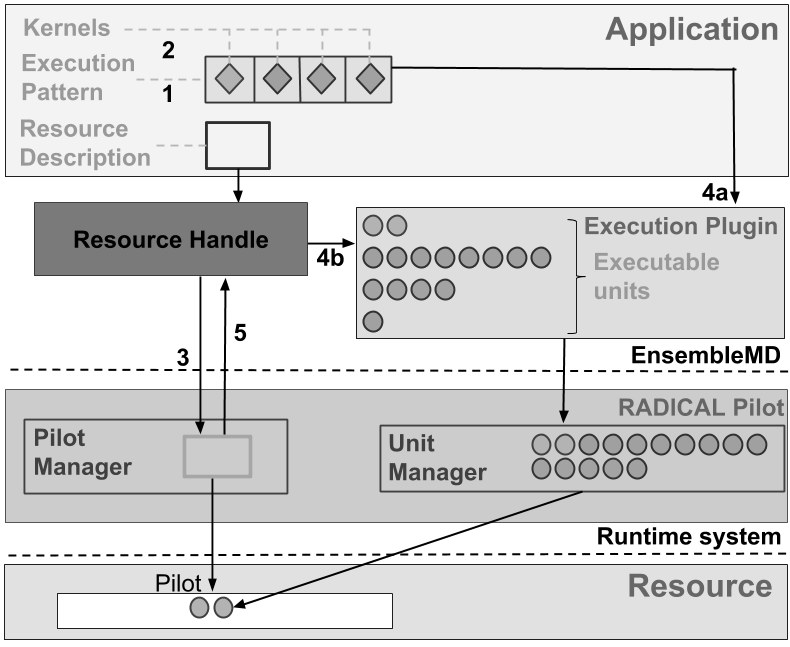}
    \caption{\enmd components: Execution Patterns, Execution Plugins, Resource Handle, Kernel plugins. Five steps: (1) Pick execution pattern for application, (2) Define Kernel plugins in the various stages of the pattern, (3) Create resource handle and submit a request for resources, (4) Execution plugin (a) binds execution pattern and the kernel plugins \& (b) executes them on the remote machine, (5) User gets back control. }
\label{fig:enmd_arch}
\end{figure}
\begin{compactenum}

\item Support the easy development of a range of ensemble-based execution patterns and applications

\item Manage execution details while providing performance

\item Support dynamic resource management so as to decouple execution details from the instantaneous availability of resources.

\end{compactenum}

% We observe that the control flow in applications across molecular sciences and other domains is quite similar. Thus, most of the applications can be categorized, based on the control flow, into a few groups.

%\jhanote{I think an execution pattern can be independent of control flow or data flow, i.e., CF or DF is how an execution pattern might be implemented. Would you agree? An execution pattern is more about ``communication pattern'' and ``synchronization pattern'' as opposed to how.. This paragraph needs attention possibly.}\mynote{fixed}

Analyzing many ensemble-based applications we observe few common and recurring ways in which ensembles communicate and are synchronized.  We take advantage of this recurrence by creating abstract ``execution patterns'' which are agnostic to the type of workload, the number of tasks and resource requirements of tasks. For example, an execution pattern of a {\it bag of tasks} would create a set of tasks that are independent of each other without specifying the actual work done or their resource requirements.

%\jhanote{I would remove the formal treatment of these and collapse into a single paragraph: With these points in consideration, we state the design decisions for \enmd:}

%Applications can then insert the specific workload.

% EnsembleMD has been designed top-down, starting from the top with only the application building blocks exposed to the user, support for the common patterns in the field of molecular dynamics down to the specifications for the execution using the MD tools. 

% Hence, \enmd itself does not provide new mechanisms for task execution and management, but relies on a runtime system with the such capabilities.

%Even with the patterns and workload specification abstractions, it is important to abstract the application specification from the complexity of workload execution, while keeping the details exposed to a minimum. Last but not least, a design goal for \enmd was to draw on existing tools, libraries and frameworks wherever possible. With these points in consideration, we state the design decisions that were made for the \enmd:

%\begin{compactenum}

%\jhanote{I would not call these out as ``bullets'' but convert the design decisions into a paragraph}\jhanote{I repeat: collapse these into a single paragraph. make them read as if they are the ``Design" as opposed to the ``design decisions/principles''. merge the following second point with the first.}

% The toolkit treats the execution pattern as a core component by providing a mechanism to develop applications by parametrizing a pre-defined patterns and defining its execution stage(s).

\jhanote{how is this different as a design feature from the above? aren't these just an instantiation of patterns?} \mynote{unit patterns are instantiations yes, but the fact we define/identify unit patterns as atomic and usable to compose higher order patterns is by design, would you not say so ?} \jhanote{if you look at layer diagram you will realize that application logic is not in scope of ensemble toolkit}\mynote{I agree, that's why I say user has to provide them, am I missing something ?}

% There could be, in principle, hundreds of execution patterns and providing parametrized templates to all of them is not practical. Hence, we

The execution pattern is a core component of the \enmd. The \enmd provides a mechanism to develop applications by parametrizing pre-defined patterns and defining its execution stage(s).  We identify and support unit patterns which are a set of atomic, unique execution patterns that capture different, exclusive modes of communication and synchronization patterns. These unit patterns can then be combined to form higher-order patterns consisting of more complex communications and synchronizations.  A user is required to provide only the application logic workload via the exposed components. The details of task creation, submission and synchronization, data management, etc., are hidden in the lower layers of \enmd while resource management, task execution, data movement, etc are performed by the underlying runtime system that \enmd utilizes.

% \jhanote{is task syncrhonization captured in runtime or the pattern, or plugin? it depends what you call the runtime; that is not defined yet}\mynote{fixed. we have, in the intro}

\shrink{
\textbf{Pattern-based approach for application development}: The toolkit treats execution patterns as a core component. \jhanote{this is a tautology: thereby capturing the synchronization and communication pattern of the application.}
The toolkit provides a mechanism to develop applications by parametrizing a pre-defined pattern and defining its execution stage(s).

\textbf{Identify and support unit patterns}: Unit patterns are a set of atomic, unique execution patterns that capture different, exclusive modes of communication and synchronization patterns. These unit patterns can then be combined to form many higher order pattern consisting of complex communications and synchronizations.  \jhanote{how is this different as a design feature from the above? aren't these just an instantiation of patterns?}

\textbf{Separation of concerns}: A user is required to provide only the application logic \jhanote{if you look at layer diagram you will realize that application logic is not in scope of ensemble toolkit} workload via the exposed components. The complex details of the task execution, task synchronization etc., are encapsulated in the runtime system. \jhanote{is task syncrhonization captured in runtime or the pattern, or plugin? it depends what you call the runtime; that is not defined yet}
}

\begin{figure*}
  \centering
   \FramedImage{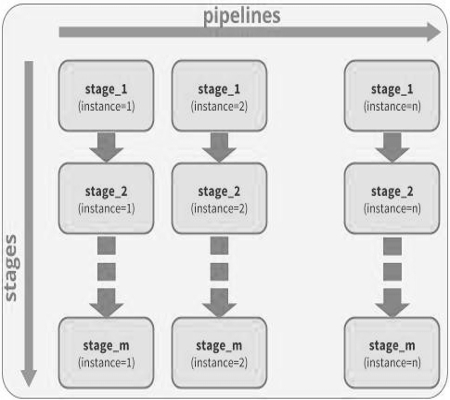}\FramedImage{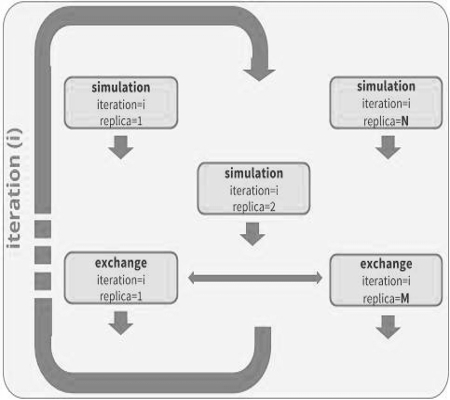}\FramedImage{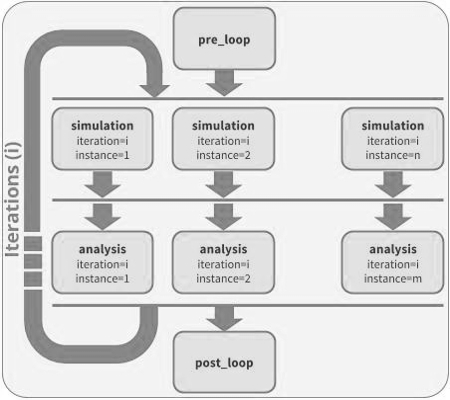}
    \caption{Execution patterns supported by \enmd : (a) Ensemble of Pipelines (b) Ensemble Exchange (c) Simulation Analysis Loop}
    \label{fig:merged_pats}
\end{figure*}

The resulting architecture of \enmd (Fig.~\ref{fig:enmd_arch}) has four components: execution patterns, resource handle, kernel plugins and execution plugins. These components ease application development while confining the execution complexity to the runtime system.  We discuss each component:

\begin{compactenum}

\item \textbf{Execution Pattern:} It is a high-level object that represents the synchronization and communication patterns of ensembles. They can be viewed as parametrized templates that capture the execution of the ensemble(s). \shrink{Execution patterns provide placeholder methods for the individual stages of an execution trajectory.}

\item \textbf{Kernel plugin}: It is an object that abstracts a computational task in \enmd. It represents an instantiation of a specific science tool along with the required software environment. Kernel plugin hides kernel-specific peculiarities across different resources as well as differences between their interfaces \shrink{ of the various kernels. to the extent possible, thus addressing kernel-level heterogeneity.}

\item \textbf{Resource Handle}: It provides methods to allocate, run an execution pattern and de-allocate resources. 

\item \textbf{Execution Plugin}: The execution plugin binds the kernel plugins and the execution pattern, and translates the tasks into executable units.  Along with resource details from the resource handle, these executable units are forwarded to the underlying runtime system, thus decoupling execution from the expression of the application. 

%\jhanote{this is future work, should not be in the definition of an existing component:} .\mynote{fixed}

% \jhanote{this maybe in the future, but currently isn't there and thus reviewers will kill us for mentioning it but providing no details: execution plugins to analyze an application's control flow, combine the results with existing information about the execution resource and optimize along various parameters: total time to completion, amount of data transferred, throughput, etc.}\mynote{yes, sorry this needs to be in future work. already there.}

\end{compactenum}

% It is important to understand how these components interact during the course of the application development. \atnote{Sorry, I don't understand what this sentence is trying to say} 

The resource handle, execution pattern and kernel plugins are user facing components used to obtain details regarding the resources, pattern used and workload at each stage of the pattern. The execution plugin is an internal component of the toolkit which is responsible for receiving information from other components and passing to the runtime system. It currently supports static binding and translation. "Intelligence" can be added so as to support applications with specific workload or resource requirements. To do so requires the use of information about the resource(s) coupled with requirements from the application, to devise execution strategies~\cite{ipdps-2016}

%\jhanote{I think the following is an implementation issue not a design issue/consideration. Shouldn't it be under implementation then?}
%\mynote{I don't think so. Regardless of how the RTS is implemented, I think this process will remain the same. }

\up

\subsection{Implementation}

%\jhanote{not good style to begin a subsection and then immediately a new subsubsection, i.e., some intervening text is good practise}

In order to keep the software footprint of \enmd small, existing tools, libraries and frameworks were used wherever possible.  To be able to use multiple HPC systems, \enmd follows a standard job submission language. 

% Although the criteria of the runtime system is to be simply compatible with this job submission language alone, we select \rp due to its specific advantages discussed below.

\subsubsection{Job submission language}
The \enmd is designed to be use the SAGA API~\cite{saga-x} which is consistent with the standard Job Submission Description Language~\cite{anjomshoaa2005job}.

%An example of the values required by the \enmd resource handle to construct the job submission script is provided in listing ~\ref{lst:res_han}.

\subsubsection{Runtime system}

%\jhanote{Is the use of Pilot system a design feature, or is the use of RP an implementation issue?}

%\jhanote{Reduce the discussion of RP. this is not a paper on RP}

\shrink{A resource placeholder, thus, decouples the acquisition of the resources from their use to execute application tasks.}

\enmd relies on a runtime system to manage task execution, data movement and resource management.  One of the design objectives was to provide dynamic resource management, a type of which is to be able to execute more tasks than available resources would allow. \enmd achieves this via the use of Pilot-Job systems~\cite{review_pilotreview}. Pilot-Job systems provide placeholders or container jobs that are submitted to the target resource. These container jobs enable application-level scheduling of any number of workload tasks on these resources.  Of the many pilot-systems currently developed \cite{review_pilotreview}, we select \rp (RP) \cite{review_radicalpilot_2015}. This is due to certain features unique to \rp: 1) Support for MPI-based applications, 2) well-characterized performance~\cite{review_radicalpilot_2015} and, 3) designed to support execution on heterogeneous resources due to its architecture and use of SAGA.  It is important to note that the \enmd can be coupled with other general-purpose or special-purpose runtime systems. \enmd is actively developed and provides the necessary hooks for this coupling.

% RP builds upon important conceptual theoretical advances~\cite{review_pilotreview}.
\shrink{Resource-level heterogeneity is addressed in \rp via an interoperability layer, SAGA-python~\cite{saga-x} to get access to diverse DCI middleware.}
%\atnote{I don't think it is fair to say that we have selected RP, since other systems are not even mentioned.}\mynote{are you suggesting we use "select" or is it something else ? aren't we "selecting" RP because of the features mentioned above. We can use \enmd with other RTS, but it might not give us this features}

\up

\subsection{Ensemble toolkit Execution Patterns}

The \enmd supports three execution patterns: ensemble of pipelines, ensemble exchange (EE) and simulation analysis loop (SAL). These patterns were motivated by their frequent use in the molecular and geo-sciences. These patterns can be used for any application that follows the same orchestration of tasks by modifying the specific execution kernel plugins.  We take a look at these patterns in terms of the coupling and dependencies of tasks.

\subsubsection{Ensemble of Pipelines}

%\jhanote{this description is using the ``ensemble of pipelines'' } \mynote{I propose we don't call it "ensemble of pipelines" in this paper since we don't discuss "pipeline of ensembles". Significance of "ensemble" in "ensemble of pipelines" doesn't show up till we discuss the other. We can use "ensemble" in its definition of course.}\jhanote{I don't understand. are you proposing to redefine the definition of a pipeline, or the pipeline pattern?}\mynote{ I meant lets use "pipeline" as the pattern name for now. although it is in fact an ensemble of pipelines}

It consists of an ensemble of independent pipelines of tasks. Each pipeline consists of multiple stages representing a well-defined execution order; each stage can contain heterogeneous workloads.  Although each stage of a pipeline depends on its predecessor, the pipelines execute independent of each other.  Figure~\ref{fig:merged_pats}(a) is a pictorial representation of the
ensemble of pipelines pattern consisting of \texttt{N} pipelines each with \texttt{M} stages. 

\shrink{ Many applications in bioinformatics follow the pipeline pattern\cite{martin2010rnnotator} \cite{ragothaman2014developing}.}

% \jhanote{I don't think tying the EE pattern to pipeline pattern is very useful. For two reasons: one it violates the general notion of a pipeline.. secondly, each pattern in principle is supposed to be conceptually independent like a linear space, so that complex workflows can be constructed out of them.}\mynote{I agree, I believe the EE pattern is a variant of the pipeline pattern. For the purpose of this paper, I think we can keep them softly independent.}

% \jhanote{you are confusing concurrency with synchronization. The way it reads currently gives the impression that the exchange and simulation are always globally synchronized}\mynote{took a crack at it}

% \jhanote{I don't think there should be an iteration! It could be that the exchanges are between ensemble members/replicas that do not ever exchange again, i.e., the exchange pattern is not iterative!}\mynote{fixed}

\subsubsection{Ensemble Exchange (EE)} The EE pattern consists of interacting ensemble members, where an ensemble member can be in one of two possible states.  An ensemble member may interact with other ensemble members while in the same stage. There is no obligatory global barrier (synchronization) across ensemble members. As depicted in Figure~\ref{fig:merged_pats}(b), each ensemble member executes independently but interacts with other members in the exchange stage.  A common example of the EE pattern is the replica exchange molecular dynamics (REMD) algorithm where the exchange between the MD simulations is not temporally synchronized and is pairwise.\shrink{\cite{sugita1999replica}.} 

% \jhanote{Needs some attention: I think there is some conflation between sequence -- a logical order, and a pipeline -- a realization of the sequence}.\mynote{fixed}

\subsubsection{Simulation Analysis Loop (SAL)}

% \jhanote{is there a hard synchronization between all the simulations before they enter the analysis?}\mynote{yes} \jhanote{by ``Set'' you mean ``ensemble''??} \mynote{yup, fixed}

\jhanote{the following description needs improvement:}\mynote{fixed} The SAL pattern is a two-stage iterative execution pattern. The first stage consists of  an ensemble of simulation instances and second stage consists of an ensemble of analysis instances. Figure~\ref{fig:merged_pats}(c) represents the SAL pattern with N simulation instances followed by M analysis instances in each iteration.  Simulation ensemble members are synchronized before transitioning, as are the analysis ensemble members.  

\shrink{This pattern finds use in many MD \cite{preto2014fast, laughton2009coco} and geoscience applications \cite{lattner2012ensemble, galmarini2004ensemble, cervone2008risk, galmarini2001forecasting, mallet2009ozone, hamill2010future,wan2014short}}

\section{Characterization and Validation}
\label{sec:exps}

\jhanote{Vivek: Heavy edits. Please review for any errors} \mynote{looks good} In this section, we characterize the performance of the toolkit and validate its design.  We measure the the total time to completion (TTC) but decompose it into relevant primary components.  We divide our experiments into two tracks: validation and performance characterization. The execution patterns are characterized by using the same workload for three patterns. Next, we examine the effect of switching kernels while the pattern is kept constant.  For the second experiments track, we characterize the scalability of \enmd -- we run strong and weak scaling tests for the EE and SAL execution patterns. 

%\subsection{Experiment Setup}

The HPC machines used for the experiments are the same platforms which scientific applications that use \enmd will use for production runs. In the first two experiments, we use the XSEDE Comet \cite{xsede_resources}, an Intel Xeon cluster with 1984 nodes and 24 cores per node and 120GB memory per node. In the scaling experiments, we use XSEDE Stampede \cite{xsede_resources}, an Intel Xeon cluster with 6400 nodes and 16 cores per node and 32GB memory per node and XSEDE Supermic \cite{xsede_supermic}, an Intel Xeon Phi cluster with 360 nodes with 20 cores per node and 60GB memory per node.

\shrink{\textbf{Workload description: } In the first experiment, we use a toy application containing 2 stages: file creation with random characters, character count of these files. In the second experiment, we construct the DM-d-MD application \cite{preto2014fast} using the SAL pattern. The physical system used was a decalanine residue with each simulation run for 1ps using the Gromacs MD engine. Next, we run scaling experiments for the SAL pattern using a solvated alanine dipeptide molecule with 2881 atoms and the Amber MD engine. }

%\atnote{Why not having a separate workload description for each set of experiments? I find this paragraph very confusing.}\mynote{maybe we can reiterate it in individual exp section, but I think the introduction should talk about it too. Do you think itemizing it helps ?}

%for 0.6 ps. For the Ensemble Exchange In the next stage, we use the CoCo analysis tool to generate a new set of coordinates.

\up
\subsection{Characterization of Execution Patterns}

To validate three existing execution patterns in \enmd -- ensemble of pipelines, EE, and SAL, we create a two-stage application for each pattern. We use the \texttt{mkfile} kernel in the first stage to create a file in each task and the \texttt{ccount} kernel in the second stage to count the number of characters in each of those files. 

%Listings ~\ref{lst:expA_pipeline} and ~\ref{lst:expA_sal} represent how the code would look while using the pipeline and SAL patterns.

In this experiment, we vary the number of tasks and cores from 24-192, keeping their ratio 1:1. We use the XSEDE allocated Comet cluster. A decomposition of the total time using the three different patterns is given in Fig ~\ref{fig:exp_setA_merged}. The first three subplots depict the execution time of the application using the three patterns. The following subplot presents the decomposition of the \enmd overhead for the ensemble of pipelines pattern. As expected, we observed similar overheads when using EE and SAL execution patterns, hence we avoid presenting them again.

%\begin{compactitem}

From the first three subplots, it can be observed that the application execution times remain relatively similar at all the configurations across patterns. This is due to the fact that each task has the same workload and all the tasks execute concurrently in all the patterns.

The Core overhead, which is the time taken to initialize \enmd, launch and cancel resources requests, remains constant in all the configurations as it is independent of the pattern or resource. The \enmd Pattern overhead is the time taken by \enmd in creating tasks using \rp \jhanote{what does constructing mean?}\mynote{fixed} and submitting tasks for execution to RP.  As expected it depends on the number of tasks. \jhanote{How does it depend on the number of tasks?}\mynote{what do you mean - linearly, quadratically ?} \jhanote{What about EnTK initialization. That should factor into core overhead? }\mynote{yes, fixed} \shrink{In subplots 4 and 5, we present the behaviour of the overheads when using the pipeline pattern.}  \jhanote{There is a major problem: The y-axis on Plot 1,2 and 3 are less than the y-axis on plot 4. An explanation is that the workloads are different. Either way, this an error as overheads cannot be greater than the TTC!}\mynote{those were TTE not TTC, the plots were ok. Anyway plot 5,6 are now removed}.  \jhanote{OK lets keep plot 4 but remove plot 5 and 6. Plot 5 is already invalid.}  There are overheads from RP (not shown) that arise from the time for task submission by RP on the remote machine, communication latencies, etc~\cite{review_radicalpilot_2015}.  \rp is being engineered to minimize these delays.

\up

\jhanote{the previous three sentences wont make sense to a reader as they don't anything about RP}\mynote{I added some detail. Do you want me to remove this part or explain RP in detail ?}\jhanote{You are losing the opportunity to convey the essential point: (i) the primary contribution to the core overhead is from RP,
(ii) RP is being engineered to lower the overhead}\mynote{mentioned above}

%\shrink{{The data in each task is \jhanote{moved and transferred? fix}\mynote{thanks, fixed} transferred from the local machine to XSEDE Comet in this experiment. Hence, in the last subplot, we see that the data movement time increases with increase in the number of tasks. This is because the amount of data increases with increase in the number of tasks. \jhanote{what is the point of this experiment??? to measure network speed? just because you were asked to do an experiment to check for consistency and errors does not mean it makes it into the paper. Experiments are performed for building insight, validation, testing hypothesis... not all of which make it into a paper}}\mynote{not required anymore since we have removed the subplot}

\jhanote{I would remove subplot 4-6 and describe them more clearly in words. very difficult to follow figures as well as weak description possibly due to space limitations} \mynote{I would definitely keep subplot 4, ok with removing subplot 6. Let's rephrase the description as required.} \jhanote{can you justify plot 4 please?
Also my suggestion was 4-6, ie., 4, 5 and 6, and not just 4 and 6.}

\shrink{It is important to note that the patterns themselves do not influence the workload. \jhanote{it is unclear what the previous sentence means?} As is evident from the results, given the exact same workload, we observe similar execution times across the different patterns.\jhanote{if the patterns are doing different things why is this the case. unclear at best}}
%\end{compactitem}

%\jhanote{so how does this validate the patterns?}

\subsection{Validation of Kernel Plugins}

The objective of this experiment is to present and validate the support for kernel abstractions in \enmd. We choose the SAL pattern and replace the kernel plugins with MD tools: Gromacs in the simulation stage and LSDMap \cite{preto2014fast} in the analysis stage. We use Comet and keep the data points \jhanote{scales of what??}\mynote{fixed} in the same range (24-192).

\begin{figure}
\centering
    \includegraphics[width=8.5cm,height=6cm]{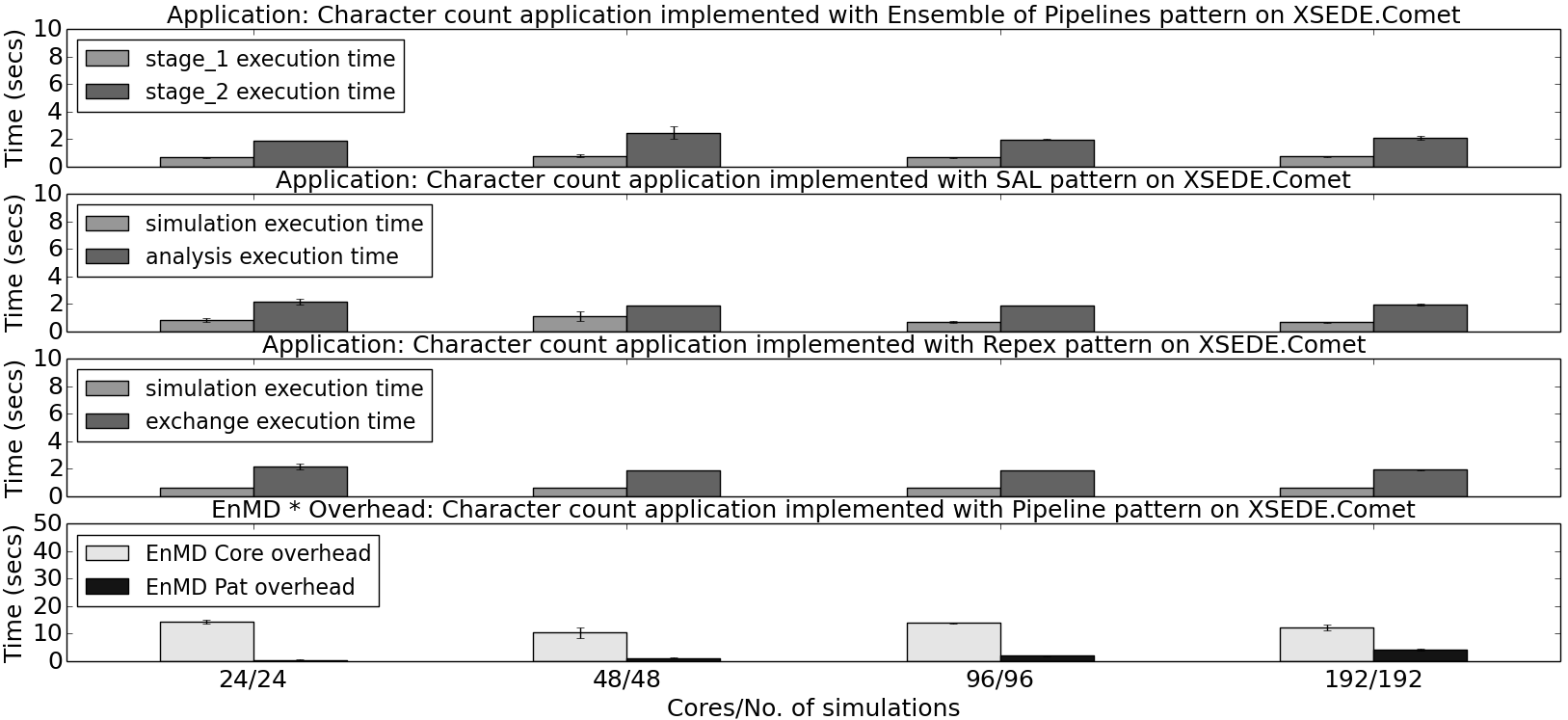}
    \caption{Character count application implemented with pipeline, SAL and EE pattern on the XSEDE Comet. We vary, in same proportion, the number of tasks and cores from 24-192, thus all the tasks are concurrently executed. The first three subplots show the application execution times using the three patterns. The next subplot shows the \enmd overhead. \jhanote{the figure has step; the text has stage}\mynote{fixed}
    }
    \label{fig:exp_setA_merged}
\end{figure}

\begin{figure}
\centering
	\includegraphics[width=8.5cm,keepaspectratio]{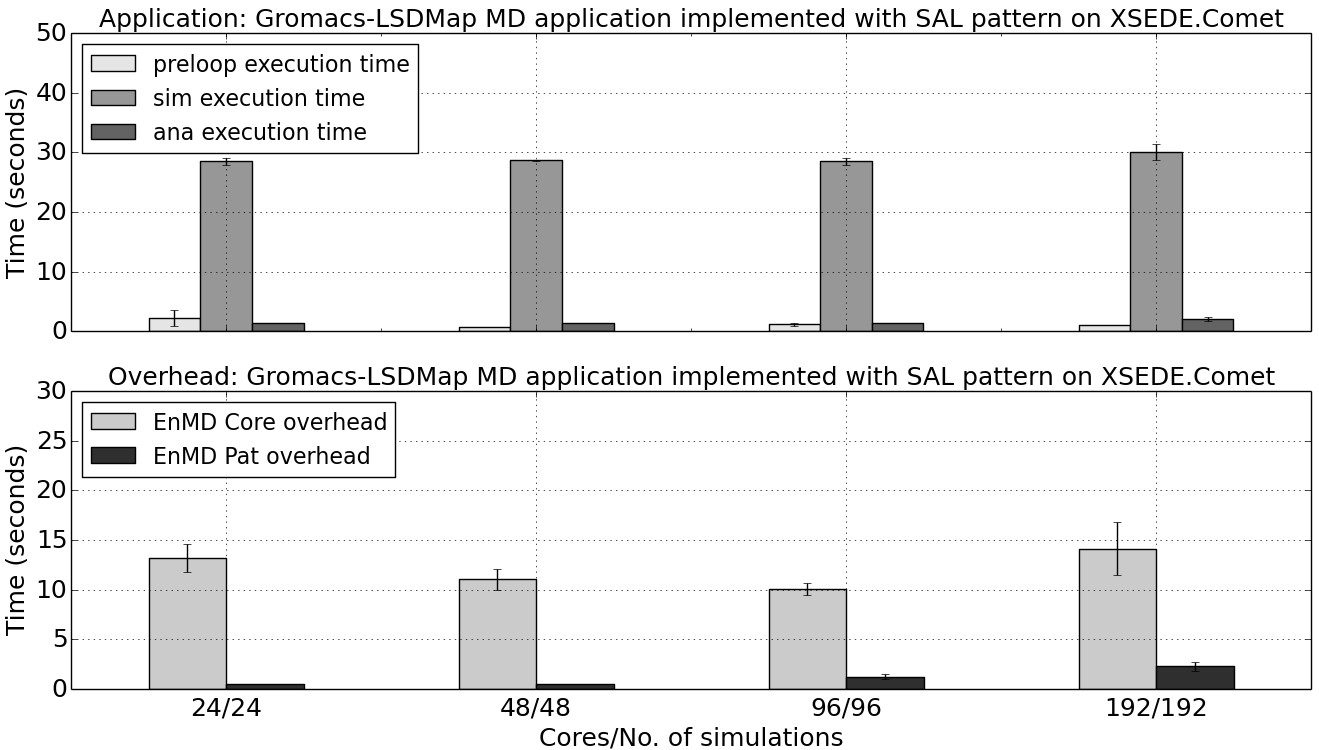}
	\caption{Gromacs-LSDMap application implemented with SAL execution pattern on the Comet cluster in XSEDE. We vary the number of tasks from 24-192 but also vary the number of cores similarly. 
%\jhanote{need to run these for the same range/values as for the Amber-COCO pattern to compare the overhead of different kernel plugins but same pattern}
        }
	\label{fig:exp_setB_sal_grlsd_comet}
\end{figure}

Figure~\ref{fig:exp_setB_sal_grlsd_comet} presents a decomposition of the total time. % We skip the data movement time since it is a factor of the amount of data and the network between the client and remote
We observe that, for the same range of tasks on the same machine, the overheads obtained with Gromacs and LSDMap are similar to the ones presented in Figure~\ref{fig:exp_setA_merged}. It can be inferred that changing the kernel plugins and hence the workload does not effect the overhead presented by \enmd.

\shrink{ Due to change in workload, now have a preloop stage and different values for the application execution time.}

\subsection{Characterization of Scalability}

We have now validated both the support for different execution patterns and kernel plugins in \enmd. We now test the scalability of the toolkit with real science workloads. We perform strong and weak scaling tests for EE pattern and the SAL pattern.

\subsubsection{Ensemble Exchange Pattern}

\begin{figure}
\centering
    \includegraphics[width=8.5cm,keepaspectratio]{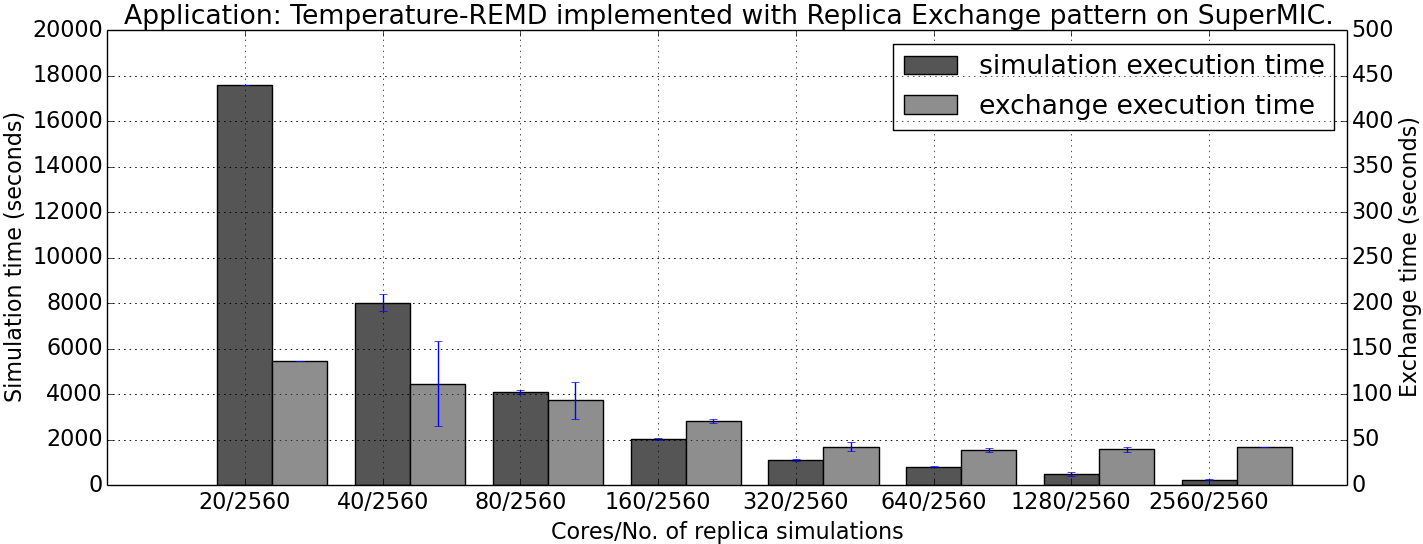}
    \caption{Strong scaling test for EE execution pattern on
          XSEDE Supermic using Amber-Temperature Exchange kernel plugins and the
          alanine dipeptide molecule with 2560 replicas over a (20-2560) range of
          number of cores hence varying problem size per core.
          }
    \label{fig:exp_setC_repex_strong}
\end{figure}

\begin{figure}
\centering
    \includegraphics[width=8.5cm,keepaspectratio]{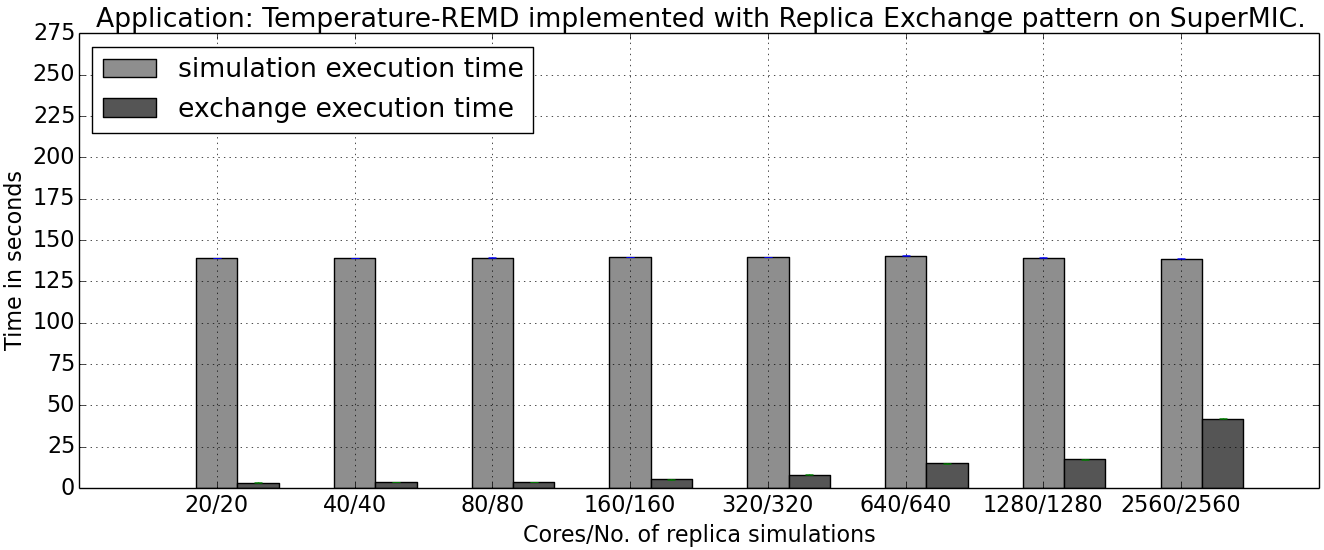}
    \caption{Weak scaling test for EE execution pattern  on XSEDE Supermic using Amber-Temperature Exchange kernel plugins and the alanine dipeptide molecule with fixed problem size per core. %
    }
    \label{fig:exp_setC_repex_weak}
\end{figure}

We run the EE pattern for a solvated alanine dipeptide molecule containing 2881 atoms. We use the Amber MD Engine for the simulations and perform a temperature exchange during the exchange stage. We perform both the experiments on the SuperMIC cluster in XSEDE \cite{xsede_supermic}. Figure~\ref{fig:exp_setC_repex_strong} and Figure~\ref{fig:exp_setC_repex_weak} present the results of the strong and weak scaling experiments respectively.

For strong scaling experiments, we keep the number of replicas constant at 2560 and vary the number of cores between 20-2560. Each replica is run on 1 core for 6ps before exchange. In Figure~\ref{fig:exp_setC_repex_strong}, note that since the simulation and exchange times are different by orders of magnitude, we use two y-axes as labeled. From Figure~\ref{fig:exp_setC_repex_strong}, we can observe that the simulation time decreases to half its value when the number of cores are doubled.  The exchange times, on the other hand, remain constant as they depend on the number of replicas, which is constant for this experiment.

In weak scaling experiments, we keep problem size per core constant, i.e., we keep the ratio of the number of replicas to the number of cores constant. We vary the number of replicas from 20-2560 and the number of cores proportionately. Each replica is run on 1 core for 6ps before the exchange. Our results in Figure~\ref{fig:exp_setC_repex_weak} show that the simulation time remains relatively constant. The exchange times, however, increases as this depends on the number of replicas.

\subsubsection{Simulation Analysis Loop Pattern}

\begin{figure}
\centering
\includegraphics[width=8.5cm,keepaspectratio]{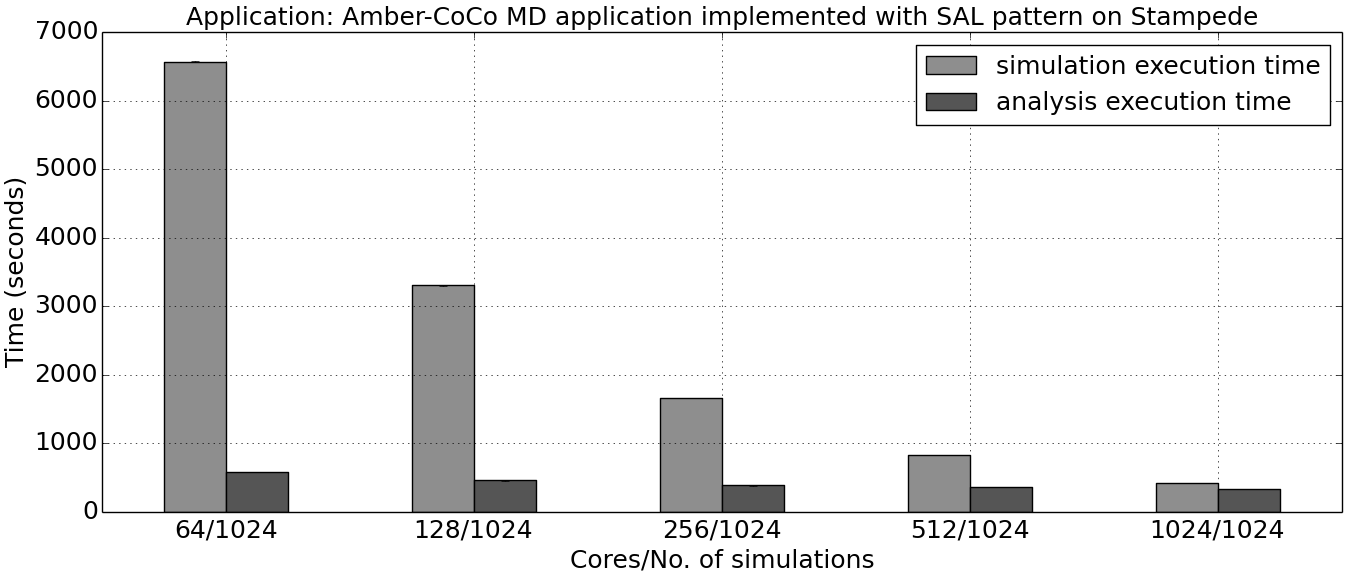}
	\caption{Strong scaling test for Simulation-Analysis-Loop execution pattern on XSEDE Stampede using Amber-CoCo\cite{laughton2009coco} kernel plugins and the alanine dipeptide molecule with 1024 simulations over a (64-1024) range of number of cores hence varying problem size per core. % There are three subplots: (i) Measurements of the overhead decomposed into components (ii) Measurements of the execution time of the application decomposed into the three stages, (iii) Measurements of the data movement time. \jhanote{Why is there no pre-loop bar in (a)? Either way needs mention}
        }
	\label{fig:exp_setC_sal_strong}
\end{figure}

\begin{figure} 
\centering
	\includegraphics[width=8.5cm,keepaspectratio]{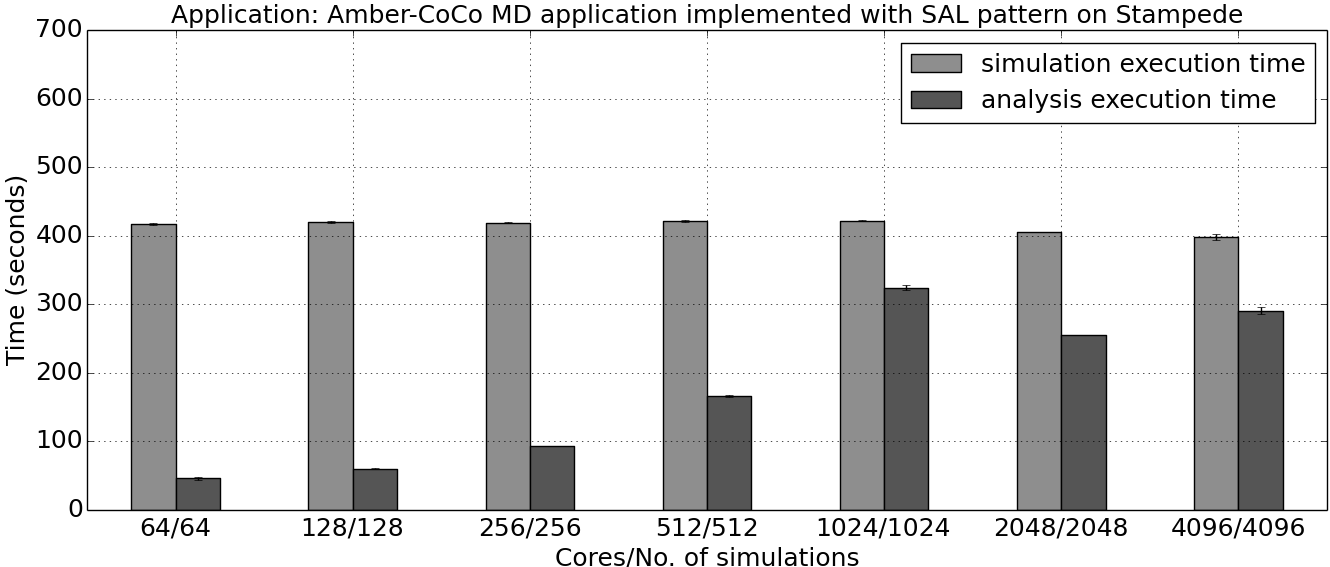}
	\caption{Weak scaling test for Simulation-Analysis-Loop execution pattern on XSEDE Stampede using Amber-CoCo kernel plugins and the alanine dipeptide molecule with number of replicas equal to the number of cores hence fixed problem size per core.% There are three subplots: (i) Measurements of the overhead decomposed into components (ii) Measurements of the execution time of the application decomposed into the three stages, (iii) Measurements of the data movement time
	}
	\label{fig:exp_setC_sal_weak}
\end{figure}

We now characterize the scalability of the toolkit by performing strong and weak scaling tests for the SAL pattern.  We implement the iterative collective coordinates algorithm\cite{laughton2009coco} using the SAL pattern. We use a solvated alanine dipeptide molecule containing 2881 atoms as our physical system. Each simulation is executed using the Amber MD Engine for 0.6 ps followed by the CoCo analysis of all simulations. Figure ~\ref{fig:exp_setC_sal_strong} and Figure~\ref{fig:exp_setC_sal_weak} present the results of our strong and weak scaling experiments.

In the strong scaling experiment, we keep the number of simulations fixed at 1024 and use one core per simulation. We vary the number of cores used from 64-1024. We observe that the simulation time decreases linearly with increase in the number of cores. The analysis algorithm is executed in serial and thus depends on the number of simulations. Hence, the analysis execution time remains constant for all configurations.

% \jhanote{there is a difference between active and passive voice, and what is foregrounded:``foregrounding research instead of the researchers''}\mynote{fixed}

In the weak scaling experiments, the number of cores is varied between 64-4096 while the number of simulations is kept equal to the number of cores at all times. For each of these configurations, the simulation execution time is observed to be constant. The analysis is executed in serial and thus increases with the number of simulations. Note that the absolute performance of the analysis kernel is unrelated to the scalability of \enmd.

% \jhanote{So what is it that our experiments tell us? see comments at the end of EE pattern.. those apply here too. and then some remarks about scalability being invariant with the pattern...}\mynote{next subsubsection}

\atnote{For experiments there is a lot of repetition explaining weak and strong scaling concept. I would assume that the reader already should be familiar with a concept.}

%\subsubsection{Analysis of scaling results}

% \jhanote{Are the results consistent with expectation, or not? Why?} \jhanote{How much further can weak/strong scaling go? What are the limits to weak/strong scaling?}\mynote{attempted}\mynote{done, please check}

\jhanote{do you mean concurrently?}\mynote{no in parallel, concurrently would be when multiple tasks t1 and t2 are executed on the same core but interleaved, t1 for a while and then t2 then back to t1.} \jhanote{lets discuss the difference between parallel and concurrent} \jhanote{the previous two sentences contradict each other!}\mynote{no, I don't think so. Total number of simulations is constant. Number of simulations executing in parallel increase with increase in core count}

\textbf{Analysis of scaling results:} In the strong scaling experiments, as we increase the number of cores, the number of simulations executing in parallel increase. As the total number of simulations is greater than or equal to the number of cores for any configuration, the TTC decreases with increasing core availability.  In the weak scaling experiments, we keep the problem size per core constant and use as many cores as there are simulations. Thus, all simulations execute in parallel at all configurations \shrink{(of cores and simulations)} and we observe almost constant execution time. \shrink{ The analysis stage in the SAL pattern execute in serial and thus dependent on the number of replicas. As expected, their execution time increases with increase in the number of simulations.}

The observed linear behaviour suggests that scalability is invariant of the patterns. The linear scaling with the size of the ensemble, i.e., number of ensemble members, is a consequence of the capabilities of the runtime system. This behaviour attests to both an important feature of \enmd and validates the choice of using \rp as the runtime system.

% \jhanote{you must point out or show data that shows that the above scaling characteristics is not a consequence of replicas being a single core. the same would hold true for replicas that are > 1 core, in fact arbitrary size. as the overheads etc are determined mostly by replica count and not replica size. Need to discuss}\mynote{made an attempt}

It important to note that although we have used one core for each simulation in these experiments, running multi-core simulations will not change the scaling behaviour. This is due to the fact the overheads, both from \enmd and \rp , depend on the number of tasks as opposed to the size of each task.  \shrink{\rp is extensible to support most multi-core execution modes~\cite{review_radicalpilot_2015}.} Whereas the size of each task might affect the absolute execution time, the behaviour across different configurations of cores and tasks would remain consistent with the results obtained for single-core simulations.

\jhanote{can you point to a graph somewhere? in extasy paper?}\mynote{we have a graph for this paper (in the 10 page version), I removed it due to space constraints. do we make room for a graph + atleast 5 lines for its explanation ? how should we proceed ?} \jhanote{As discussed, we need to reference the arxiv paper in this paper}

\subsubsection{MPI capability}

\begin{figure}
\centering
    \includegraphics[width=8.5cm,keepaspectratio]{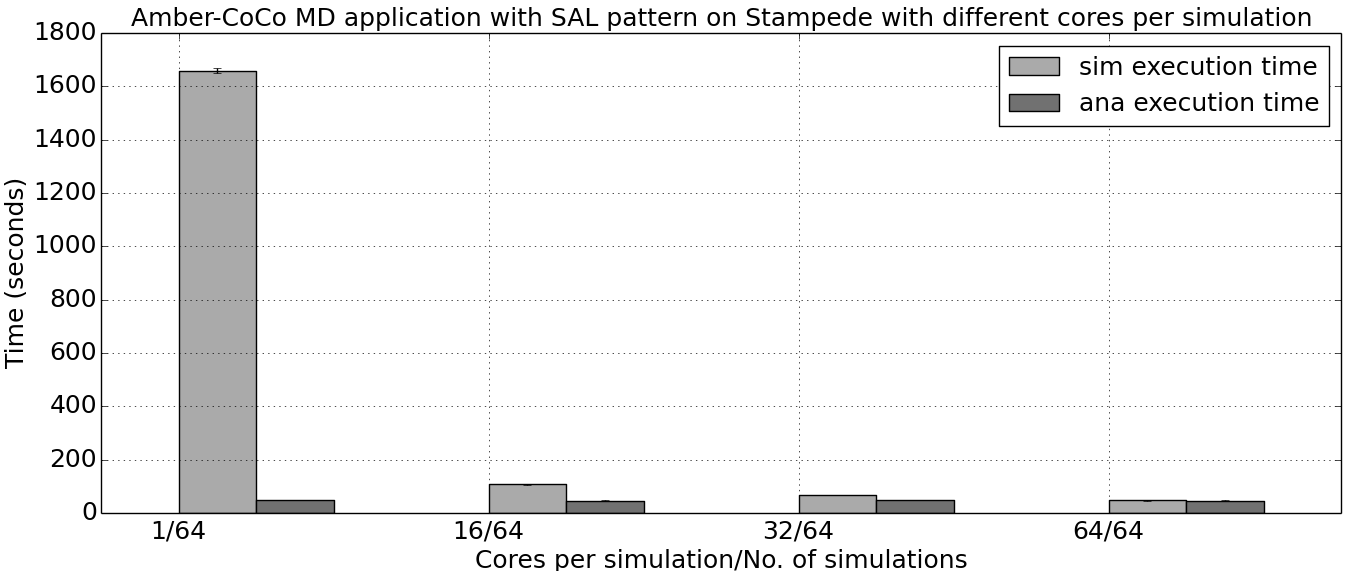}
\caption{Amber-CoCo MD application using the SAL execution pattern on XSEDE Stampede with varying no. of cores per simulation for a fixed no. of simulations}
\label{fig:exp_sal_mpi}
\end{figure}

We briefly demonstrate the MPI support in \enmd by using the same Amber-CoCo MD application, but now with more than one core per simulation. In Figure~\ref{fig:exp_sal_mpi}, we fix the number of concurrent simulations at 64, increase the duration of simulation ten-folds to 6ps and vary the number of cores per simulation as 1,16,32,64 (and hence the total no. of cores 64, 1024, 2048, 4096). We see that the execution time of the simulations drops linearly with the number of cores used. Along with the strong and weak scaling experiments, this attests to the fact that given access to the required amount of resources, O(1000) tasks of both MPI and non-MPI type can be supported by \enmd.

%Section 2 offered a description of the design objectives, toolkit design and a glance of the supported patterns and the respective user interface. Section three discusses the three sets of experiments aimed at validating the different components of the toolkit. In this section, we discuss how we believe we have addressed the design objectives set in section 2 with the experiments laid out in section three. We conclude this section with the constraints or limitations that exist with the current incarnation of the toolkit and how we aim to address the same.

\section{Discussion and Conclusion}

The objective of \enmd is to construct simple yet flexible approaches to compose ensemble-based workflows.  In Section~\ref{sec:enmdtk} we outlined its primary requirements, design and implementation.  Execution patterns expose common coordination and synchronization patterns, while the user provides workload definition and resource information. We introduced three execution patterns that represent the majority of ensemble-based applications and serve as useful unit patterns from which to construct more complex patterns.  \shrink{Section~\ref{sec:enmdtk}, concluded with a discussion of the interface of the user-facing components and an end-to-end example, which served to demonstrate the separation of concerns (ease of development and flexibility versus execution).} Experiments in Section~\ref{sec:exps}, validated the design of execution patterns and kernel plugins and characterized the performance of \enmd as a function of ensemble size.

Scalability experiments for the EE and SAL pattern with real workload showed linear strong and weak scaling behavior for up to O(1000) tasks. As measured by overheads, \enmd does not impose any significant or fundamental scalability limits; performance and scalability are essentially determined by the pilot-based runtime. \rp has been engineered to support up to 8K tasks on XSEDE Stampede and 2K tasks on Cray machines~\cite{review_radicalpilot_2015,rp_cug_2016}.  O(10,000) tasks are being tested currently on NSF Blue Waters machines~\cite{pracaward}.  A technical road-map exists for O(100,000) concurrent tasks. These runtime performance enhancements will be seamlessly applicable to \enmd.

% We characterize the toolkit in Section~\ref{sec:exps} to validate the support for different execution patterns and kernel plugins to show that both are agnostic to the actual workload. We perform s

%support of multiple execution patterns and kernel plugins. We also 

% This generic approach is powered by kernel plugins to specify the application workload.

% The customization of ensemble execution pattern templates, however, provides the simplicity of a black-box approach, while providing the ability to compose ensemble-based workflows. 

% different ways to compose applications but it is not known a priori which is likely to be most 

\jhanote{Vivek: please read carefully and fix}\mynote{looks good} This paper provides two important lessons for the workflow community: First, that the ``building blocks'' approach to constructing ``functionally-constrained'' or ``special-purpose'' workflow systems as exemplified by \enmd is a promising one. Second, special-purpose workflow systems are likely to be more useful for complex but structured workflows such as the bio-molecular applications that adhere to the patterns in this paper.  In other words, special-purpose workflow systems such as \enmd provide an effective middle ground between re-factoring applications to use generic but difficult workflow systems, and completely unstructured approaches such as scripting. 

\shrink{ \enmd provides building blocks for developing and executing an ensemble-based application.} \shrink{which is critical as there is a need to support a range of requirements without wholesale refactoring.}

\shrink{Some tools however, may choose to expose fewer degrees-of-freedom and customization options.  For example, ExTASY (\url{http://extasy-project.org/}) -- which can be viewed as a domain specific workflow tool, uses \enmd, but limits the type of execution patterns supported.  \enmd thus provides some empirical data in the ``schism'' and debate between general-purpose but monolithic workflow system versus building blocks and abstractions based approach to constructing functionally specialized tools to support special-purpose workflow. }

\enmd is currently in prototype stage, but it supports several active ensemble-based \shrink{science} applications as part of the ExTASY~\cite{extasy-2016} project and replica-exchange applications~\cite{review_repex_2016}. These applications expose fewer degrees-of-freedom and customization options but are functionally complete and provide required performance. \enmd thus provides some empirical data in the ``schism'' and debate between general-purpose but monolithic workflow system versus special-purpose workflows constructed using building blocks and abstraction-based approaches.

There are multiple enhancements to \enmd that are planned or underway: One avenue of research is to identify a {\it complete set} of unit patterns that can be used to compose {\it any} higher order ``complex'' pattern. The ability to express these higher order patterns as functions of unit patterns will enhance the ability to support more applications. \shrink{For example, applications involving seismic inversion~\cite{tape2007finite}, atmospheric forecasting of air quality, ozone forecast, modelling hurricanes, risk from nuclear release are extensions of existing unit patterns.}

Currently, workloads are adapted to resources that are chosen based upon user choice and independent of the dynamic state of workloads. Ref.~\cite{ipdps-2016} formalized and introduced {\it execution strategies} as the time-ordered set of decisions needed to execute a workload on dynamic resources. This requires integrating both application-level and resource information, which will see the execution plug-in transition from being a simple translation layer to an intelligent middleware component.  The transition from static workload-resource mapping to dynamic mapping will enable efficient and optimized execution capabilities. It will lead to the ability to efficiently select resources for a given workload, as well as form the basis to adapt workloads to optimally utilize a pre-specified set of resources.

% \jhanote{Unclear what this sentence is trying to say:} \mynote{reworked. That the complexity is not only in execution of these workloads but also that the user should be able to define what the decision points are and what path is to be taken from each decision point. this requires additional user-facing methods and possibly rework of existing user-facing methods} 

Many applications are not pre-defined as the type and amount of workload may be determined during execution time. \enmd will progressively support more adaptive scenarios, for example the ability to kill-replace tasks\cite{rapidaward}, vary the number of tasks between stages, vary the workload in each task during execution time.  The complexities arise from the adaptive execution and workloads, as well as the challenge in providing user-facing components to convey the decision points, and thus the adaptivity, meaningfully. In general, \enmd forms an initial prototype of a software system upon which to develop advanced adaptive simulation algorithms~\cite{nextgen-molsim}.
%\shrink{, some initial ideas of which are sketched in~\cite{nextgen-molsim}.}  
\jhanote{Please be consistent with the style/format for references, i.e., space after the cite}\mynote{sorry. ok.}

{\bf Software and Data:} \enmd can be found at \url{http://radicalensemblemd.readthedocs.org/en/latest} and is released under the MIT license. An updated version of this paper is available at \url{http://arxiv.org/abs/1602.00678}. Raw data and scripts to reproduce experiments can be found at \url{https://github.com/radical-experiments/enmd-pattern-testing}.

{\footnotesize {\bf Acknowledgements: } We thank members of the RADICAL group for significant discussion in the design, testing and documentation of \enmd: Ming Tai Ha for his feedback and work on testing and documentation, Mark Santcroos and Andre Merzky for help with RADICAL-Pilot, Nikhil Shenoy for initial experiments. We also thank Iain Bethune (EPCC) for testing and feedback on software, and other members of the ExTASY project. We also thank Peter Kasson, Thomas Cheatham and Michael Shirts for useful discussion about adaptive execution patterns. We also thank Matthieu Lefebvre, Ryan Modrak and Jeroen Tromp for insight into ensemble applications in seismic tomography. This work was funded by NSF CHE-1265788 and NSF ACI 1440677.}

\bibliographystyle{IEEEtran}
\bibliography{./enmd,./radical_publications}

\end{document}